\documentclass[conference]{IEEEtran}
\IEEEoverridecommandlockouts
\usepackage{cite}
\usepackage{amsmath,amssymb,amsfonts}
\usepackage{algorithmic}
\usepackage{graphicx}
\usepackage{textcomp}
\usepackage{xcolor}
\usepackage{multirow}
\usepackage{multicol}
\usepackage{booktabs}
\usepackage{svg}
\usepackage{fancyhdr}
\usepackage{hyperref}
\usepackage{url}  

\fancypagestyle{plain}{
  \fancyhf{}
  \fancyhead[C]{Conference on \LaTeX}     
  \fancyfoot[L]{This is a notice}

}
\usepackage{eso-pic}

\def\BibTeX{{\rm B\kern-.05em{\sc i\kern-.025em b}\kern-.08em
    T\kern-.1667em\lower.7ex\hbox{E}\kern-.125emX}}
\begin{document}

\title{AI Enabled High Frame Rate Portable Ultrasound Imaging Pipeline: Prototype Implementation with GPU Acceleration*\\
\thanks{The authors would like to acknowledge the support from Texas Instruments for providing the ultrasound front end and the NVIDIA Corporation for the donation of CLARA AGX Developer Kit through the Academic Hardware Grant program.}
}

\author{\IEEEauthorblockN{Arun Kumar V}
\IEEEauthorblockA{
\textit{Indian Institute of Technology Palakkad}\\
India \\
nurakumarv@gmail.com}
\and
\IEEEauthorblockN{Madhavanunni A. N.}
\IEEEauthorblockA{
\textit{Indian Institute of Technology Palakkad}\\
India \\
121813001@smail.iitpkd.ac.in}
\and
\IEEEauthorblockN{Mahesh Raveendranatha Panicker}
\IEEEauthorblockA{
\textit{Singapore Institute of Technology}\\
Singapore \\
mahesh.panicker@singaporetech.edu.sg}}

\begin{titlepage}
    \vspace*{\fill}
\fontsize{15}{18}\selectfont\textcopyright { 2024 IEEE. Personal use of this material is permitted. Permission from IEEE must be obtained for all other uses, in any current or future media, including reprinting/republishing this material for advertising or promotional purposes, creating new collective works, for resale or redistribution to servers or lists, or reuse of any copyrighted component of this work in other works.}
    \vspace*{\fill}
\end{titlepage}

\AddToShipoutPictureBG{%
  \AtPageUpperLeft{%
    \setlength\unitlength{1in}%
    \hspace*{\dimexpr0.5\paperwidth\relax}
    \makebox(0,-0.75)[c]{{This work has been accepted in the IEEE Ultrasonics, Ferroelectrics, and Frequency Control Joint Symposium 2024 (IEEE UFFC-JS 2024).}}%

}}

\AddToShipoutPictureBG{%
  \AtPageLowerLeft{%
    \setlength\unitlength{1in}%
    \hspace*{\dimexpr0.5\paperwidth\relax}
    \makebox(0,0.75)[c]{\textcolor{red}{This is an originally submitted version and has not been reviewed by independent peers.}}%
}}


\maketitle

\begin{abstract}
In this paper, we present a GPU-accelerated prototype implementation of a portable ultrasound imaging pipeline on an Nvidia CLARA AGX development kit. The raw data is acquired with nonsteered plane wave transmit using a programmable handheld open platform that supports 128-channel transmit and 64-channel receive. The received signals are transferred to the Nvidia CLARA AGX developer platform through a host system for accelerated imaging. GPU-accelerated implementation of the conventional delay and sum (DAS) beamformer along with two adaptive nonlinear beamformers and two Fourier-based techniques was performed. The feasibility of the complete pipeline and its imaging performance was evaluated with \textit{in-vitro} phantom imaging experiments and the efficacy is demonstrated with preliminary \textit{in-vivo} scans. The image quality quantified by the standard contrast and resolution metrics was comparable with that of the CPU implementation. The execution speed of the implemented beamformers was also investigated for different sizes of imaging grids and a significant speedup as high as 180 times that of the CPU implementation was observed. Since the proposed pipeline involves Nvidia CLARA AGX, there is always the potential for easy incorporation of online/active learning approaches. 

\end{abstract}


\section{Introduction}
Ultrasound (US) imaging is a popular diagnostic imaging modality because of its ease of use, affordability, real-time imaging capabilities, and lack of ionizing radiations. It is the most affordable and effective non-invasive bedside tool for soft-tissue examinations, guided interventions, and investigation of blood flow dynamics at high frame rates \cite{Tanter2014UltrafastUltrasound, lee2020point}. Driven by the recent advancements in system-on-chip devices, recent research on point-of-care ultrasound (POCUS) imaging suggests that US imaging would be the stand-out modality for ultraportable systems even at low resource settings \cite{palmer2019wireless, xu2020programmable}.



In point-of-care applications, handheld ultrasound devices have been gaining popularity, and a variety of FPGA-based handheld ultrasound devices have been proposed\cite{kim2012single, lee2014new,kim2017smart, ahn2015smartphone}. G.-D. Kim et. al. used a single FPGA (Spartan 3, Xilinx Inc.) for the portable ultrasound system. This 16-channel configuration delivers image quality comparable to that of a 32-channel ultrasound system \cite{kim2012single}, achieved through the extended aperture (EA) technique with a maximum frame rate of $30\ fps$. However, for handheld POCUS, the system's applications are limited because of its size ($245\ mm \times 190\ mm$) and weight ($560\ g$), and compatibility is poor because the system uses dedicated equipment for display. A tablet PC-based handheld POCUS has been proposed by Y. Lee et. al. using a single Xilinx Spartan FPGA \cite{lee2014new}. The system employs 16 channels, which can be extended to 32 channels using the EA method, with a maximum frame rate of $22\ fps$. It uses an 11.6-inch Tablet PC for display. In \cite{kim2017smart}, a smartphone-based POCUS that includes a 128-element transducer, a 32-channel scanner module, and a single Xilinx Artix 7 FPGA was proposed. This system uses a commercially available smartphone for backend processing with the GPU and displays the results with an interactive graphical user interface (GUI). The proposed system achieves a maximum frame rate of $50\ fps$. S. Ahn et. al. also proposed a smartphone-based POCUS using a low-cost Xilinx Spartan 6 FPGA\cite{ahn2015smartphone}, similar to the work in\cite{lee2014new}. The work also proposes a 16-channel system that can be extended to 32 channels using the EA method with a maximum frame rate of $58\ fps$.\\
In all of the approaches described above, the objective was to have a focused transmit low frame rate POCUS system. However with high frame rate ultrasound systems employing non-steered plane wave transmits are getting popular, a novel pipeline for the same is presented in this work. The proposed framework incorporates 1) high frame rate plane wave transmit 2) GPU (Nvidia CLARA AGX) accelerated beamforming algorithms (beyond the typical delay and sum beamforming as shown in most of the previous works) and 3) the potential capability to incorporate active learning for various tasks due to the availability of a 24 GB GPU with the Nvidia CLARA AGX.     
\section{Materials and Methods}
\subsection{Proposed Imaging pipeline}
\begin{figure*}[htbp]
\centering
\centerline{\includegraphics[width=0.9\textwidth]{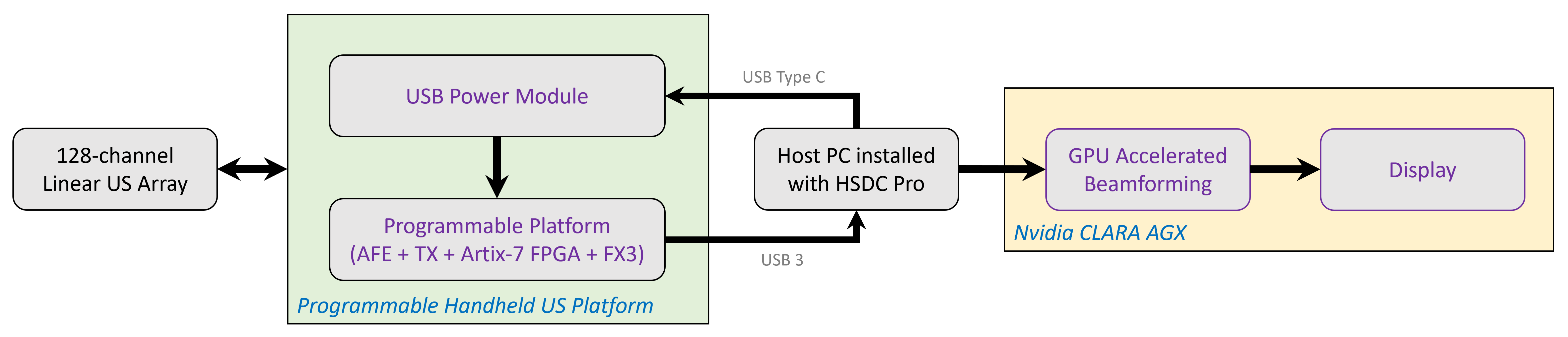}}
\caption{Proposed portable imaging pipeline.}
\label{fig:pipeline}
\end{figure*}
\begin{figure*}[t]
\centering
\centerline{\includegraphics[width=0.9\textwidth]{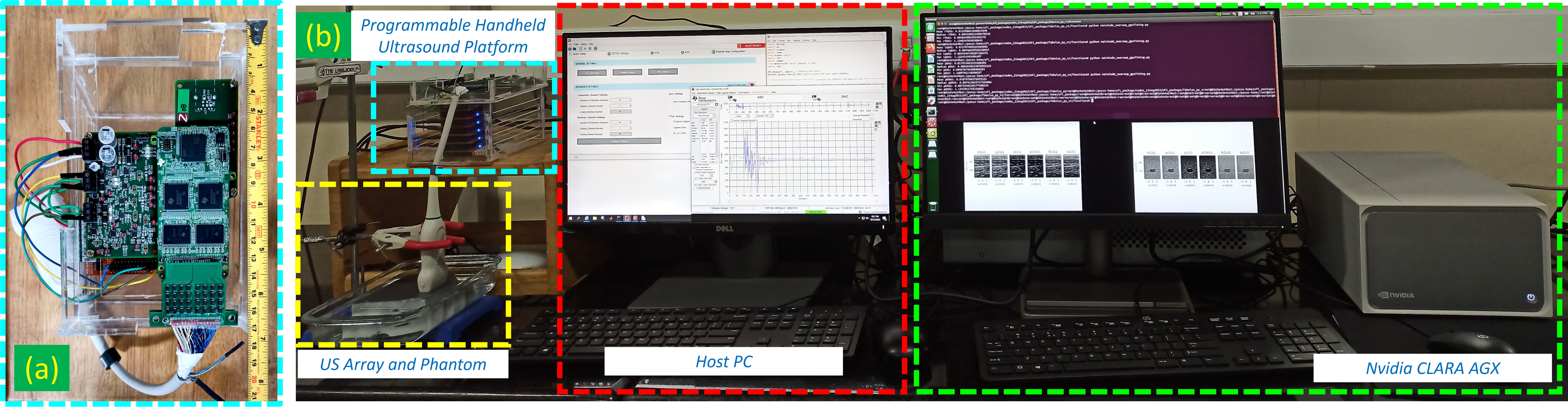}}
\caption{(a) Programmable US platform and the linear array used for the study (b) Experimental setup illustrating the complete pipeline. Each block is marked and labeled for better comprehension of the pipeline shown in Fig. \ref{fig:pipeline}.}
\label{fig:expt_setup}
\end{figure*}
The proposed GPU-accelerated portable imaging pipeline, shown in Fig. \ref{fig:pipeline}, comprises a US probe, programmable handheld US platform, host PC, and Nvidia CLARA AGX. The programmable universal serial bus (USB) power module ($65\ mm \times 53\ mm \times 63\ mm$), powered through high-speed USB Type-C ports, generates the required pulsing voltages (up to $\pm 80\ V$) and other low voltages required for the open US platform. The programmable open US platform reported in \cite{xu2020programmable} having a footprint of $150\ mm \times 100\ mm$ is used in this work to realize the proposed pipeline. It has four 32-channel transmitters and two highly integrated low-power 32-channel ultrasound Analog Front End (AFE) to support 128-channel transmit and 64-channel receive. The RF data received by the AFEs is transferred to a host computer, through a field-programmable gate array (FPGA) by the Cypress FX3 module. The FX3 provides high-speed radio frequency (RF) data transfer (throughput of up to $5\ Gb/s$) alongside functions as the USB controller that handles communication between FPGA and the host computer (host PC). The received data is further pipelined to Nvidia CLARA AGX for accelerated imaging. More details of the data acquisition and the GPU acceleration platforms are given in Table \ref{tbl:systemDetails}.

\begin{table}[t]
\centering
\caption{Details of the Data Acquisition and GPU Acceleration Platforms}\label{tbl:systemDetails}
\begin{tabular*}{0.45\textwidth}{ll}
\hline \hline 
Parameter         & Values / Details   \\
\hline
\multicolumn{2}{l}{Ultrasound Transducer Array}\\
\hline
Transducer Type         & Linear (ALS L12-5A)   \\
No. of elements ($N$)     & 128                   \\
Element pitch ($p$)           & 0.3 mm                \\
Center frequency  ($f_c$)      & 7.5 MHz               \\
\hline
\multicolumn{2}{l}{Programmable US data acquisition platform}\\
\hline
Transmitter (TX)  & 4 $\times$ TX7332\cite{tiWebTX}                \\                
Analog Front End (AFE)  & 2 $\times$ AFE5832LP\cite{tiWebAFE}               \\      
Sampling frequency ($f_s$)  & 50 MHz                \\  
ADC bit precision  & 10 bit                \\  
FPGA    &  Xilinx Artix-7 XC7A100T\cite{webFPGA} \\
USB Controller  & Cypress FX3 CYUSB301X\cite{webFX3} \\
Average Power  & $<5\ W$ \\
\hline
\multicolumn{2}{l}{GPU-accelerated beamformer implementation platform}\\
\hline
System  & Nvidia CLARA AGX\cite{webClara}                \\     
\multirow{2}{*}{GPU}  & NVIDIA RTX™ 6000      \\   
 &(4608-core NVIDIA Turing™ GPU) \\
GPU Memory  & 24GB GDDR6 with ECC                \\  
CPU  & 8-core Carmel ARM® v8.2 64-bit CPU \\  
CPU Memory    &  32GB 256-Bit LPDDR4x, 136.5GB/s\\
\hline \hline
\end{tabular*}
\end{table}

\subsection{Receive Beamforming Methods}
Conventional DAS beamformer along with two adaptive nonlinear beamformers (DAS with p-th root compression of signals (p-DAS) and delay multiply and sum (DMAS) with p-th root compression of signals (BB-DMAS)) and two Fourier-based techniques (Lu’s and Stolt’s f-k migration) were considered for this study. A brief description of the above beamformers is provided in the following subsections.

\subsubsection{Conventional Delay and Sum (DAS) Beamforming}
For an $N$-channel receive, conventional DAS beamformed signal for a pixel, $p$ at $\left[x_p,\ z_p\right]$ is expressed as:

\begin{equation}\label{eq:DAS}
    y_{DAS}[p] = \sum_{i\ =\ 1}^{N}{w_i\left[p\right] u_i[\Delta n_p]} = \boldmath{W}^H\boldmath{U}
\end{equation}

where $u_i[n]$ is the received signal at the $i^{th}$ channel, $\Delta n_p = \frac{2f_s}{c}\sqrt{(x_i-x_p)^2 + (z_i-z_p)^2}$ is the total geometric propagation delay of the US signal for $p^{th}$ pixel for a speed of sound of $c$. $w_i\left[p\right]$ and $\boldmath{W}$ are the predefined apodization weight and weight vector respectively for the pixel-element pair while $\boldmath{U}$ is the vector of delay compensated signals and $H$ denotes the Hermitian operator. In this work, DAS implementation was performed on radio-frequency (RF) and inphase and quadrature (IQ) signals which are denoted as RF-DAS and IQ-DAS respectively. A more detailed discussion on the DAS beamforming is available in \cite{perrot2021so}.

\subsubsection{$p$DAS and BB-DMAS Beamforming}
The $p$DAS \cite{polichetti2018nonlinear} and BB-DMAS \cite{shen2019ultrasound} are two nonlinear data-adaptive methods based on $p^{th}$ root compression of the RF and IQ  signals respectively. In $p$DAS proposed in \cite{polichetti2018nonlinear}, the amplitude of the delay compensated RF signal in (\ref{eq:DAS}) is compressed using $p^{th}$ root before summing it up and the dimensionality is restored by applying a signed $p$-power post summation as expressed in (\ref{eq:pDAS_a}) and (\ref{eq:pDAS_b}). 
\begin{equation}\label{eq:pDAS_a}
 \Tilde{y}_{pDAS}[p] = \sum_{i\ =\ 1}^{N}{w_i\left[p\right] sign(u_i[\Delta n_p]) |u_i[\Delta n_p]|^{1/p}}
\end{equation}
\begin{equation}\label{eq:pDAS_b}
 y_{pDAS}[p] = sign(\Tilde{y}_{pDAS}[p])\ |\Tilde{y}_{pDAS}[p]|^{p}
\end{equation}

In contrast, BB-DMAS performs $p^{th}$ root compression of delay-compensated IQ signals before summing and applies a signed $p$-power post summation without disturbing the phase as shown in (\ref{eq:BBDMAS}) \cite{shen2019ultrasound}:
\begin{equation}\label{eq:BBDMAS}
 y_{BB-DMAS}[p] = \left(\frac{1}{N}\sum_{i\ =\ 1}^{N}{\hat{s_i}[p]}\right) ^2 
\end{equation}

where $\hat{s_i} = \sqrt{a_i}\ e^{j\phi_i}$ is the delay compensated IQ signal for the $p^{th}$ pixel. The beamformed image is obtained by taking the absolute value of the $y_{BB-DMAS}$.

\subsubsection{Lu’s and Stolt’s f-k migrations}
The advanced Fourier migration techniques like Lu’s \cite{cheng2006extended} and Stolt’s f-k \cite{garcia2013stolt} beamformers reconstruct the image in Fourier or frequency-wavenumber ($f-k$) domain to reduce the compute load. Lu’s migration performs the frequency remapping of the 2D fast Fourier transform (FFT) of the raw data based on a scattering model \cite{garcia2013stolt}. In contrast, Stolt's f-k migration performs the frequency remapping based on a reflector model \cite{cheng2006extended}.
The details and differences in the frequency remapping of both methods can be seen in \cite{garcia2013stolt}. GPU-accelerated Python versions of these methods were developed from the MATLAB codes openly available at \cite{webAdrienGitlab, besson2016sparse, webGithubRAli}

All the above beamformers were implemented in Python with GPU acceleration using the Cupy library, which is an open-source NumPy- and SciPy-compatible library developed for GPU-accelerated computing \cite{webCupy}.

\subsection{Experimental Setup}
The feasibility of the proposed pipeline, imaging performance, and the execution time of the GPU-accelerated beamformers was evaluated with \textit{in-vitro} experiments. Poly-vinyl alcohol (PVA) based ultrasound phantoms with cysts and points targets were custom prepared by following the protocol reported in \cite{mercado2018effect} and were used for \textit{in-vitro} investigations. The portable US platform was configured for 64-channel nonsteered plane wave transmit and 64-channel receive for a depth of $3\ cm$. The raw RF data sampled at $50\ MHz$ was transferred to the Nvidia CLARA AGX kit for the performance evaluation of beamformers. The experimental setup is shown in Fig. \ref{fig:expt_setup}.

\subsection{Evaluation Metrics}
The imaging performance of the proposed pipeline for various beamformers is evaluated with a single nonsteered plane wave data in terms of full-width half maxima (FWHM) and the traditional contrast metrics. The FWHM is measured at $-6$ dB of the lateral and axial profiles of point targets. The contrast ratio (CR) and contrast-to-noise ratio (CNR) are measured for anechoic cyst targets using the following expressions: 
\begin{equation}
    CR\ =\ 20{log}_{10}\frac{\mu_{in}}{\mu_{out}} 
\end{equation}

\begin{equation}
    CNR\ =\ \frac{\mu_{in}-\mu_{out}}{\sqrt{\sigma_{in}^2+\sigma_{out}^2/2}}
\end{equation}
where, $\mu$ and $\sigma$ indicate the mean and standard deviation, respectively, while the subscripts {$in$} and {$out$} indicate the inside and outside of the cyst. 

The execution time of the beamformers is evaluated as the median value of the processing time taken for the reconstruction of 100 images and compared with that of the execution speed of its CPU implementation. Five different sizes of imaging grid viz. $128 \times 32$, $256 \times 32$, $512 \times 64$, $1024 \times 64$, and $1750 \times 64$ were used to evaluate the execution speed of the GPU-accelerated beamformers. The Python-based CPU implementation was performed on an Intel(R) Xeon(R) W-2155 3.30 GHz processor installed with a memory of 32 GB.

\section{Results and Discussion}
In this section, the quantitative and qualitative results from the experimental studies are presented.

\subsection{Imaging performance}
\begin{figure*}[!ht]
\centerline{\includegraphics[width = 0.95\linewidth]{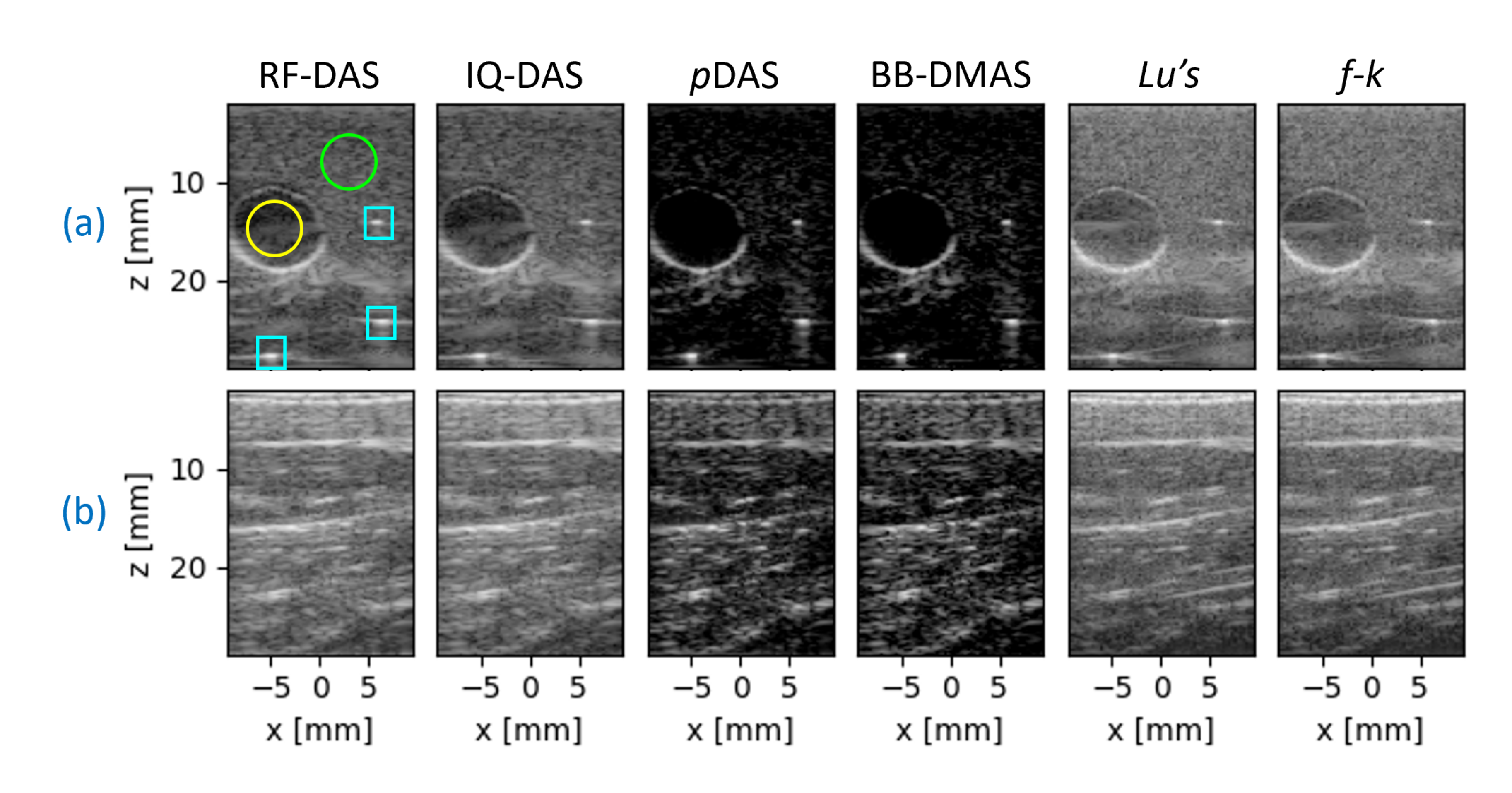}}
\caption{Experimental Results: (a) Reconstructed image of an \textit{in-vitro} phantom having anechoic cyst and point targets. (b) Reconstructed images from a sample \textit{in-vivo} scan from brachioradialis of a healthy subject. All images are shown with a dynamic range of -70 dB.}
\label{fig:cystImg}
\end{figure*}

Fig. \ref{fig:cystImg} shows the beamformed images obtained for the \textit{in-vitro} phantom experiment and a sample scan of \textit{in-vivo} brachioradialis using various beamformers. The image quality obtained with the proposed pipeline was comparable with that obtained with CPU implementation irrespective of the beamforming technique. It is evident from the contrast and resolution metrics plot shown in Fig. \ref{fig:quantitativePlots}(a). The regions marked in yellow and green circles in Fig. \ref{fig:cystImg}(a) were considered for contrast evaluation and the point targets selected for resolution measurement are shown in cyan squares in Fig. \ref{fig:cystImg}(a).

\subsection{Execution times of the beamformers}
\begin{figure*}[!htbp]
\centerline{\includegraphics[width = 0.99\linewidth]{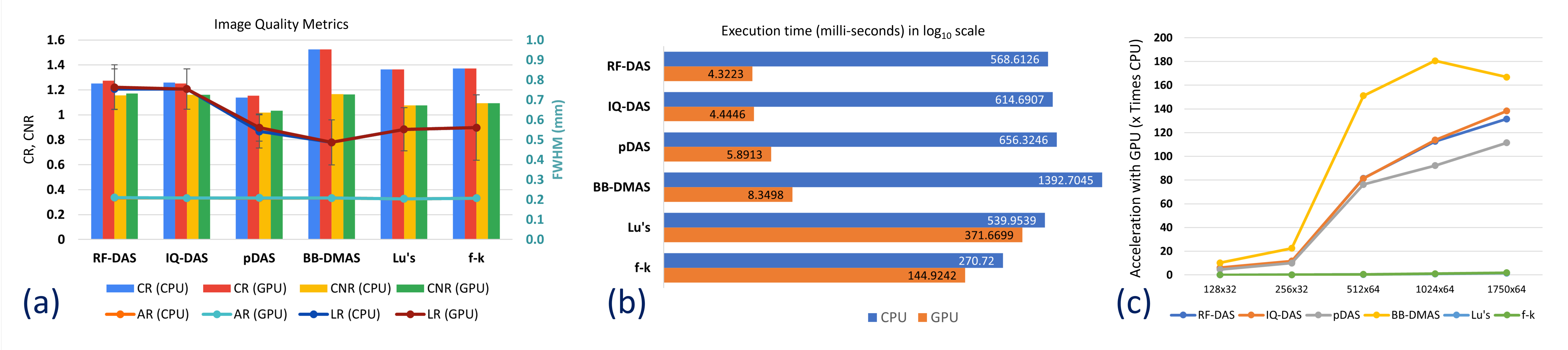}}
\caption{(a) Comparison plot of contrast and resolution metrics for CPU and GPU implementation. (b) Comparison between the execution time of various beamformers implemented in host CPU and GPU in Nvidia CLARA AGX. The time axis is in log$_{10}$ scale and the actual time is indicated in each bar in milli-seconds. (c) GPU acceleration plotted as xTimes to that of CPU counterpart for different grid sizes.}
\label{fig:quantitativePlots}
\end{figure*}

A comparison plot of the execution time of the beamformers in GPU and CPU for a grid size of $1750 \times 64$ is shown in Fig. \ref{fig:quantitativePlots}(b) and the GPU acceleration observed for various sizes of imaging grids is shown as xTimes to that of CPU counterpart in Fig. \ref{fig:quantitativePlots}(c).
A significant speedup as high as 180 times that of the CPU implementation was observed for larger grids while a speed up as low as around 5 times was observed for smaller grids with RF-DAS, IQ-DAS, pDAS, and BB-DMAS beamformers. However, Lu's and f-k migration have shown the least acceleration with GPU which needs to be looked upon.

\subsection{Limitations and Future Scope}
Even though the image quality has been observed similar for CPU and GPU implementations, the speedup obtained for Fourier-based methods was not significant. In this regard, future studies to improve the implementation of those methods are warranted. Also, the suitability and potential of other adaptive beamformers based on minimum variance techniques need to be investigated. 
An important aspect of the proposed framework is the possibility of learning-based beamformers for POCUS \cite{katare2022learning} which is a relatively unexplored area of research. With a potent 24GB GPU on board on the Nvidia CLARA AGX, there is a possibility of enabling active/continual learning for accelerating beamforming, image processing for segmentation, object detection, or even federated learning.
Moreover, the present prototype study has been limited to brightness mode (B-mode) imaging only. However, high frame rate plane wave transmit opens up the possibility of ultrafast blood flow imaging and it would be taken up as an immediate future work. 

\section{Conclusion}
This paper presented a prototype study of a portable hand-held ultrasound imaging platform integrated with GPU-accelerated implementation of beamformers. Various state-of-the-art beamformers were implemented with GPU acceleration as part of the pipeline and a significant speed-up was observed without any noticeable degradation in imaging quality when compared to that of its CPU counterparts. The ultraportable handheld data acquisition platform having a footprint of $150 mm \times 100 mm$ coupled with the computer capability of the Nvidia CLARA AGX makes the possibilities of learning-based beamformers wide open for POCUS systems.

\section*{Acknowledgment}

The authors would like to thank Mr. Shabbir Amjhera Wala and his team, Texas Instruments, Bengaluru, and Dr. Anish Bekkal, ELAGORi, Bengaluru, for the technical support and useful discussions. We acknowledge the Physical and Chemical Biology Lab, Indian Institute of Technology Palakkad, Kerala for providing the required facilities for the phantom development.\\
We also thank the support from the Texas Instruments for providing the ultrasound front end and the NVIDIA Corporation for the donation of the CLARA AGX Developer Kit through the Academic Hardware Grant program.

\bibliographystyle{ieeetr}

\small{
\bibliography{references.bib}
}

\end{document}